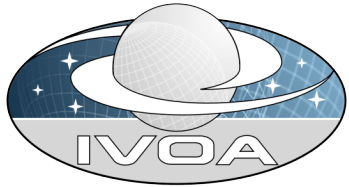

**I**nternational
**V**irtual
**O**bservatory
**A**lliance

# Resource Metadata for the Virtual Observatory Version 1.12

## IVOA Recommendation 2007 March 2

**This version:**
   http://www.ivoa.net/Documents/REC/ResMetadata/RM-20070302.html

**Latest version:**
   http://www.ivoa.net/Documents/latest/RM.html

**Previous version(s):**
   http://www.ivoa.net/Documents/PR/ResMetadata/RM-20061212.html
   http://www.ivoa.net/Documents/PR/ResMetadata/RM-20051115.html
   http://www.ivoa.net/Documents/WD/ResMetadata/RM-20050621.html
   http://www.ivoa.net/Documents/REC/ResMetadata/RM-20040426.html
   http://www.ivoa.net/Documents/PR/ResMetadata/RM-20040323.html
   http://www.ivoa.net/Documents/PR/ResMetadata/RM-20040126.html
   http://www.ivoa.net/Documents/WD/ResMetadata/RM-20031002.html
   http://www.ivoa.net/Documents/WD/ResMetadata/RM-20030801.html
   http://www.ivoa.net/Documents/WD/ResMetadata/RM-20030709.html
   http://www.ivoa.net/Documents/WD/ResMetadata/RSM-20030509.html
   http://www.ivoa.net/Documents/WD/ResMetadata/RSM-20030206.html
   http://www.ivoa.net/Documents/WD/ResMetadata/RSM-20021011.html

**Editor(s):**
   Robert Hanisch

**Author(s):**
   IVOA Resource Registry Working Group
   NVO Metadata Working Group

## Abstract
An essential capability of the Virtual Observatory is a means for describing what data and computational facilities are available where, and once identified, how to use them. The data themselves have associated metadata (e.g., FITS keywords), and similarly we require metadata about data collections and data services so that VO users can easily find information of interest.  Furthermore, such metadata are needed in order to manage distributed queries efficiently; if a user is interested in finding x-ray images there is no point in querying the HST archive, for example.  In this document we suggest an architecture for resource and service metadata and describe the relationship of this architecture to emerging Web Services standards.  We also define an initial set of metadata concepts.

## Status of this document

This is a Recommendation. The first release of this document was 7 June 2002.  This is an update to the Recommendation dated 2004 April 26.  The goal of this update is to clarify the definitions of certain metadata elements, add certain new elements, and delete elements that have not been useful.

A list of current IVOA Recommendations and other technical documents can be found at http://www.ivoa.net/Documents/.

## Acknowledgments

Many members of the IVOA Registry Working Group, AstroGrid project, NVO Technical Working Group, and participants in IVOA Interoperability workshops have made significant contributions to this document.  Contributors to this document have been partly or completely supported by the following projects and programs:

- The U.S. National Virtual Observatory project, which is funded by the National Science Foundation's Information Technology Research Program under Cooperative Agreement AST0122449 with The Johns Hopkins University.
- The UK AstroGrid project, which is funded by the Particle Physics and Astronomy Research Council.
- The Astrophysical Virtual Observatory, which is funded by the fifth framework program of the European Community for research, technological development, and demonstration activities (FP5).

## Contents

# 1  Introduction

An essential capability of the Virtual Observatory is a means for describing what data and computational facilities are available where, and once identified, how to use them. The data themselves have associated metadata (e.g., FITS keywords), and similarly we require metadata about data collections and data services so that VO users can easily find information of interest. Furthermore, such metadata are needed in order to manage distributed queries efficiently; if a user is interested in finding x-ray images there is no point in querying the HST archive, for example. In this document we suggest an architecture for resource and service metadata and describe the relationship of this architecture to emerging Web Services standards. We also define an initial set of metadata concepts.

# 2  Architecture

In order to make it easy for astronomy information services to participate in the VO, we propose a hierarchical system for metadata management. At the top level we require a minimum amount of information, sufficient primarily to note the existence of a resource and to describe who is responsible for it. At lower levels, the metadata are more extensive and complex, allowing for the description of query syntax, access protocols, and usage policies.

A *resource* is a general term referring to a VO element that can be described in terms of who curates or maintains it and which can be given a name and a unique identifier. Just about anything can be a resource: it can be an abstract idea, such as sky coverage or an instrumental setup, or it can be fairly concrete, like an organisation or a data collection. This definition is consistent with its use in the general Web community as "anything that has an identity" (Berners-Lee 1998, IETF RFC2396). We expand on this definition by saying that it is also describable.

An *organisation* is specific type of resource that brings people together to pursue participation in VO applications. Organisations can be hierarchical and range greatly in size and scope. At a high level, an organisation could be a university, observatory, or government agency. At a finer level, it could be a specific scientific project, space mission, or individual researcher. A *provider* is an organisation that makes data and/or services available to users over the network.

A *service* is any VO resource that can be invoked by the user to perform some action on their behalf. Associated with any service is descriptive *metadata* about the service. Metadata generally include information the user needs to determine if a service is of interest and how the service may be invoked. Specific types of metadata are described below. Note that the service itself need not be aware of the metadata that describe it.

A *query service* supports a query/response protocol. The user submits a query to the service that may define characteristics of interest, and the service returns a set of information to the user. The query may be null, e.g., a current-time service may only support a null query, and some services may respond to a null query with appropriate default actions. Non-query services may also exist, e.g., services to copy or delete files on remote files systems, to mail information to other users, to kill existing jobs, authorize actions, etc.

A *registry* is a query service for which the response is a structured description of resources. The resources described by a registry may be of any type. The registry may support a query that allows the user to indicate which resources might be of interest.

In our model, the hierarchy of resources is one in terms of management and curation. For example, an organisation may manage a collection of one or more services and even smaller organisations or projects. For example, MAST, HEASARC, IRSA, NED et al. are all resources. Each of these manages other resources, e.g., the HST archive in MAST. They also support specific services (which are also resources) such as an HST observation log query service or a cone search service. One could in principle describe all of NASA astrophysics data holdings as a resource, or all of NVO as a resource, but aggregates of this scale circumvent the goal of being able to locate the specific resources and services of interest for a particular application.

All resources are described by metadata. *Resource metadata* are generic, high-level, and independent of any specific service. Resource metadata include

- *Identity metadata,* which gives the resource a name and an identifier,

- *Curation metadata*, which describe who supports the resource and its availability (i.e., version, release date), and

- *Content metadata*, which describe what kind of information is available (types of data, sky coverage, spectral coverage, etc.). Content metadata can be either general, applying to all resources, or associated more specifically with data collections and the services that deliver data from them.

Resource metadata are typically not queryable parameters in the underlying services, but rather they encompass information that now is simply "known" to users, or must be discovered through other means. Astronomers know that the HST archive includes optical images and spectra, for example, or that Vizier provides access to catalogs and tables. Resource metadata constitute a "yellow pages" of astronomical information. Resource metadata are analogous to the UDDI (Universal Description, Discovery and Integration) Web Service, and are analogous to the high-level descriptions included in the CDS GLU.

Organisations, data collections, and services can be considered as classes of resources that may each require additional metadata to fully describe it, but which are not shared by other classes. For example, a service description would need to include its inputs, outputs, and how it can be accessed. *Service metadata*, therefore, can be thought of as an extension of the general resource metadata: where as the resource metadata, through its content metadata, describes *what* is available, the service metadata describes *how* to access it.

Resource metadata will be collected through resource registration services, e.g., web forms that present a resource curator with the requisite fields and enumerated lists, and construct a resource descriptor in a standard format (such as VOTable). The resource registration service should not allow fields to be left unspecified. Some metadata elements may be irrelevant, unknown, or not provided by the publisher of a resource. Since "irrelevant" conveys different information than "not provided", we will adopt standard representations of these conditions:

| | |
|---|---|
| "Not Applicable" | irrelevant or not applicable to this resource |
| "Unknown" | unknown, cannot be defined |
| "Not Provided" | no information was provided by the resource publisher |

Various applications based on the registry may choose to include or exclude certain resources based on these attributes. If a metadata element is "Not Provided" the application should make no assumption regarding applicability or relevance.

Similarly, some resources may provide quite large aggregations or collections, covering many bandpasses, types, or formats. It may be prohibitive to list all such options. In such cases acceptable representations for the metadata entries would be:

| | |
|---|---|
| "Any" | resource will respond to requests for any of the available types (though some may not actually be available) |
| "All" | resource will respond to requests for all of the available types, and all are actually available in some non-zero quantity |

The most general resource metadata is similar in concept to the Dublin Core metadata definitions (http://dublincore.org/documents/dces/), and where possible DC metadata elements have been used. VO metadata elements that correspond directly to DC counterparts are noted. The Dublin Core elements Language and Relation are not currently used in the VO metadata.

## 3   Resource metadata concepts

Below we describe the *concepts* we believe are needed in the resource metadata. These concepts may be instantiated in a variety of standard forms, e.g. XML, UCD tags, or FITS keywords, and with a variety of mechanisms, such as Topic Maps, OWL, or RDBMSs. Consequently, the exact names and rendering of the values may depend on the particular form in which they are represented. For example, when *Coverage.Spatial* is rendered as a FITS keyword record, the name will need to be limited to 8 characters and the value rendered in a pure ASCII form; in contrast, when rendered in XML, it might be better to tag the different components of the value separately. It will be necessary to define standard renderings for each of these common forms.

A limited number of keywords are considered essential for a basic understanding of the resource, and are thus denoted as *required*. All others are optional, or may be applied to certain classes of resources only.

## 3.1 Identity metadata

*Title* (string) [Dublin Core] [Required]
Definition: A name given to the resource.
Comment: Typically, a Title will be a name by which the resource is formally known. Title should be an unabbreviated form (e.g., Hubble Space Telescope) rather than an acronym unless the acronym is so well known as to be part of standard usage. Publishers are encouraged, but not required, to define unique Titles.

*ShortName* (string)
Definition: A short abbreviation for the name given to the resource.
Comment: The ShortName will be used where brief annotations for the resource name are desired, such as in GUIs that might refer to many resources in a compact display. ShortName strings are limited to a maximum of sixteen characters. Care should be taken to define illuminating ShortNames indicating either where the resource comes from or what data collection it provides. ShortNames are not required to be unique. Indeed, a resource provider may use the same ShortName for several related resources (e.g., different services that access the same collection), or the same ShortName might be used by different providers for common/mirrored resources. In the latter case, the ShortName defined by the original publisher of the resource should have preference.

*Identifier* (URI) [Dublin Core] [Required]
Definition: An unambiguous reference to the resource within a given context. The syntax for Identifiers is described in *IVOA Identifiers* in the IVOA document collection (http://www.ivoa.net/Documents/).
Comment: The URI corresponding to the resource.

## 3.2 Curation metadata

*Publisher* (string) [Dublin Core] [Required]
Definition: An entity responsible for making the resource available
Comment: Examples of a Publisher include a person or an organisation. Users of the resource should include Publisher in subsequent credits and acknowledgments.

*PublisherID* (URI)
Definition: The identifier for the entity responsible for making the resource available. The syntax for Identifiers is described in *IVOA Identifiers* in the IVOA document collection (http://www.ivoa.net/Documents/).
Comment: This item is optional; an ID for the publisher may not yet be established (e.g., if the publisher has not yet been registered).

*Creator* (string) [Dublin Core]
Definition: An entity primarily responsible for making the content of the resource.
Comment: Examples of a Creator include a person or an organisation. Users of the resource should include Creator in subsequent credits and acknowledgments. Creator is intended to refer to the organisation or individuals responsible for the intellectual content of the resource, and not the organisation or individuals who may have developed the service by which the content is made available. Guidelines: 1) If the resource is a data collection or service accessing a collection, then Creator fields should list the scientists responsible for the original data collection. Typically, this would be list of authors associated with the defining published paper for the collection. At a minimum, the PI or lead author should be given. Full names should be given, not just surnames. 2) For a collection that is a compilation of many separately published collections (e.g., an archive), then the Creator should be set to "various". 3) If the resource is an organisation not associated with a specific collection, the most appropriate value is either empty or the name of the person responsible to assembling the organisation. Often, an empty value is most appropriate. 4) If the resource is a Registry that publishes records for a single organisation, the Creator may contain the person(s) responsible for collecting or creating the metadata held in its records. Otherwise, it can be an empty value. 5) If the resource is an Authority, it should contain the name of the person that reserved the authority ID it records.

>  *Creator.Logo* (URL)
>  Definition: A URL pointing to a graphical logo, which may be used to help identify the information resource.

*Contributor* (string)     [Dublin Core]
Definition: An entity responsible for making contributions to the content of the resource.
Comment: Examples of a Contributor include a person or an organisation. Users of the resource should include Contributor in subsequent credits and acknowledgments. Like Creator, Contributor is intended to refer to the organisation or individuals responsible for the intellectual content of the resource, and not the organisation or individuals who may have developed the service by which the content is made available. Also see the Guidelines under Creator.

*Date* (string)     [Dublin Core] [Required]
Definition: A date associated with an event in the life cycle of the resource. Typically, Date will be associated with the creation or availability (i.e., most recent release or version) of the resource. ISO8601 is the preferred format (YYYY-MM-DD).
Comment: Dates may be approximate (e.g., year only, year and month). When the resource is an organisation, Date should refer to the approximate genesis of the organisation. When the resource is a service, Date should refer to the implementation date or the date the service came available. When the resource describes an authority identifier, Date should refer to when the authority identifier was reserved. (See *IVOA Identifiers* in the IVOA document collection (http://www.ivoa.net/Documents/)).

*Version* (string)
Definition: A label associated with the creation or availability (i.e., most recent release or version) of the resource.

*Contact* (string, e-mail address)
Definition: The e-mail address for contacting the persons responsible for the resource.
Comment: Contact is split into two components for clarity.

>  *Contact.Name* (string)
>  Definition: The name of the contact.

Comment:  A person's name, "John P. Jones", or a group, "Archive Support Team".

*Contact.Address* (string)
Definition:  The mailing address of the contact.
Comment:  All components of the mailing address are given in one string, e.g., "3700 San Martin Drive, Baltimore, MD 21218  USA"

*Contact.Email* (e-mail address)
Definition:  The e-mail address of the contact.
Comment:  For example, "John.P.Jones@navy.gov", or "archive@datacenter.org".

*Contact.Telephone* (string)
Definition:  The telephone number of the contact.
Comment:  Complete international dialing codes should be given, e.g., "+1-410-338-1234"

## 3.3  General content metadata

*Subject* (string, list)     [Dublin Core] [Required]
Definition:  A list of the topics, object types, or other descriptive keywords about the resource.
Comment:  Subject is intended to provide additional information about the nature of the information provided by the resource.  Is this a catalog of quasars?  Of planetary nebulae?  Is this a tool for computing ephemerides?  Terms for Subject should be drawn from the IAU Astronomy Thesaurus (http://msowww.anu.edu.au/library/thesaurus/), though in the absence of suitable terms (the IAU Thesaurus is not complete in all areas of astronomical research) the following alternate collections of astronomical research terms may be used:
   Vizier keywords (CDS):  http://vizier.u-strasbg.fr/doc/ADCkwds.htx
   Astronomy journal keywords:
http://www.aanda.org/index2.php?option=com_content&task=view&id=170&Itemid=184
Guidelines:  As this is a Required element, it must not be left blank.  Services that provide access to data from registered collections should replicate the Subject metadata in their registry entries.  To support keyword-based searches of registry contents, the Subject element should be as specific as possible and include as many relevant terms as possible.

*Description* (string, free text)  [Dublin Core] [Required]
Definition: An account of the content of the resource.
Comment:  Description may include but is not limited to: an abstract, table of contents, reference to a graphical representation of content or a free-text account of the content. Thorough text descriptions are particularly encouraged in order to make text-based searches against the registries maximally useful.  Description should emphasize what the resource is about, as other matters such as who created it, when it was created, and where it is located are described elsewhere in the resource metadata.

*Source*  (string)        [Dublin Core]

Definition: A bibliographic reference from which the present resource is derived or extracted.
Comment: The present resource may be derived from the Source in whole or in part. Recommended best practice is to use the standard *bibcode* (see http://cdsweb.u-strasbg.fr/simbad/refcode.html), where available. If no *bibcode* is available, Source should use a string or number conforming to a formal identification or citation system.

*ReferenceURL* (URL)   [Required]
Definition: A URL pointing to additional information about the resource. In general, this information should be human-readable.

*Type* (string, list)   [Dublin Core] [Required]
Definition: The nature or genre of the content of the resource.
Comment: Type includes terms describing general categories, functions, genres, or aggregation levels for content. VO Types include:

| Type | Description |
| --- | --- |
| Archive | Collection of pointed observations |
| Bibliography | Collection of bibliographic references, abstracts, and publications |
| Catalog | Collection of derived data, primarily in tabular form |
| Journal | Collection of scholarly publications under common editorial policy |
| Library | Collection of published materials (journals, books, etc.) |
| Simulation | Theoretical simulation or model |
| Survey | Collection of observations covering substantial and contiguous areas of the sky |
| Education | Collection of materials appropriate for educational use, such as teaching resources, curricula, etc. |
| Outreach | Collection of materials appropriate for public outreach, such as press releases and photo galleries |
| EPOResource | Collection of materials that may be suitable for EPO products but which are not in final product form, as in Type Outreach or Type Education. EPOResource would apply, e.g., to archives with easily accessed preview images or to surveys with easy-to-use images. |
| Animation | Animation clips of astronomical phenomena |
| Artwork | Artists' renderings of astronomical phenomena or objects |
| Background | Background information on astronomical phenomena or objects |
| BasicData | Compilations of basic astronomical facts about objects, such as approximate distance or membership in constellation. |
| Historical | Historical information about astronomical objects. |
| Photographic | Publication-quality photographs of astronomical objects. |
| Press | Press releases about astronomical objects. |
| Organisation | An organisation that is a publisher or curator of other resources. |
| Project | A project that is a publisher or curator of other resources. |
| Registry | A query service for which the response is a structured description of resources. |
| Other | A resource not described by any of the above types. |

This list is extensible. Resources providing more than one type of content should list all relevant types.

*ContentLevel* (string, list)
Definition: A description of the content level, or intended audience.
Comment: VO resources will be available to professional astronomers, amateur astronomers, educators, and the general public. These different audiences need a way to find material appropriate for their needs.

| ContentLevel | Definition |
| --- | --- |
| General | Resource provides information appropriate for all users |
| Elementary Education | Resource provides information appropriate for grades K-4 education |
| Middle School Education | Resource provides information appropriate for grades 5-8 education |
| Secondary Education | Resource provides information appropriate for grades 9-12 education |
| Community College | Resource provides information appropriate for education at community colleges |
| University | Resource provides information appropriate for university-level education |
| Research | Resource provides information appropriate for professional-level research and graduate school education |
| Amateur | Resource provides information of interest to amateur astronomers |
| Informal Education | Resource provides information appropriate for education at museums, planetariums, and other centers of informal learning |

*Relationship* (string)
Definition: A resource may be related to another resource in a way that is important to document, so that associated services or duplicate copies may easily be located.

| | |
| --- | --- |
| mirror-of | The resource is a mirror of another resource. Information gathered from the resources is indistinguishable. |
| service-for | The resource is a service associated with a data collection. |
| derived-from | The resource is a derivative of another resource, e.g., a subset selected for a particular scientific interest, or a reprocessed data collection. |
| served-by | The resource (e.g., a data collection) can be accessed via another service resource. |

*RelationshipID* (URI)
Definition: The identifier of an associated resource. The relationship is described in the Relationship metadata element. The syntax for Identifiers is described in *IVOA Identifiers* in the IVOA document collection (http://www.ivoa.net/Documents/).

## 3.4 Collection and service content metadata

*Facility* (string, list)
Definition: The observatory or facility where the data was obtained.
Comments: Some resources are likely to hold data from multiple observatories. If just a few, this could be a list; if very many, just say "many". Theoretical data will not originate with an observatory, but rather might be characterized by the computational facility used to create them (NCSA, SDSC, etc.).
Comments: Facility should be used only to describe entities that specifically produce or manage data. Observatory names are the most common values. When the resource is an organisation, Facility may include the names of archives or well known services (e.g. NED) that one may obtain data from. The listing of Facility values need not be complete; rather, it can be indicative of the facilities that are most important or of most common interest. The value may be "various" when many facilities are associated with the resource. The value may be empty when there is no facility that is particularly relevant to the resource.

*Instrument* (string, list)
Definition: The instrument used to collect the data.
Comments: Can be a specific instrument name (Wide Field/Planetary Camera 2) or generic instrument type (CCD camera). Theoretical data is produced by a computer code, and the name of the code could be specified.

*Coverage* (string)     [Dublin Core, with modifications]
Definition: The extent of scope of the content of the resource.
Comment: The Dublin Core notion of coverage is too generic to be of much use in the VO, where we need more specific information. We propose to subset this element as follows:

> *Coverage.Spatial* (string)
> Definition: The sky coverage of the resource.
> Comment: The complete syntax for the spatial coverage specification is described in the Space-Time Coordinate (STC) Metadata definition document ResourceProfile/AstroCoordArea definition (http://www.ivoa.net/Documents/latest/STC.html). Resource metadata may be somewhat simplified (i.e., do not give detailed spatial coverage of a large archive), but should be expressed in a syntax which adheres to the STC specification. All positions should be given in degrees.

> | Region Name | Specification |
> |---|---|
> | Circle | Circle [fillfactor *<fill>*] *<frame>* *<ξ>* *<η>* *<radius>* |
> | Polygon | Polygon [fillfactor *<fill>*] *<frame>* *<ξ$_1$>* *<η$_1$>* ... |
> | Box | Box [fillfactor *<fill>*] *<frame>* *<ξ>* *<η>* *<ξ$_{size}$>* *<η$_{size}$>* |
> | PositionInterval | PositionInterval [fillfactor *<fill>*] *<frame>* *<ξ$_{min}$>* *<η$_{min}$>* *<ξ$_{max}$>* *<η$_{max}$>* |
> | AllSky | AllSky [fillfactor *<fill>*] *<frame>* |

> In the above,

- *<frame>* is one of the following: ICRS, FK5, FK4, J2000, B1950, ECLIPTIC, GALACTIC[_II], SUPER_GALACTIC, GEO_C, GEO_D, UNKNOWN; the default value is UNKNOWN.
- *<ξ> <η>* are generic longitude and latitude coordinates in the given *<frame>*, such as right ascension and declination or galactic longitude and latitude, given in floating point degrees
- *<radius>* is a floating point value, in degrees
- [...] indicates that the preceding parameter set may occur more than once, to specify a "support" area that consists of multiple intervals.
- *<fill>* (single interval only) is a number between 0 and 1 indicating the fraction of the region covered; the default value is 1.0.

Unions of disjoint or overlapping regions are indicated by a simple concatenation of region specification strings. The sides of Polygons and Boxes are assumed to be segments of great circles. PositionInterval can be used to define regions limited by a minimum and maximum declination (small circles).

*Coverage.RegionOfRegard* (float, decimal degrees)
Definition: RegionOfRegard is a single numeric value representing the angle by which a positional query against this resource should be "blurred" in order to get an appropriate match.

In the case of image repositories, it might refer to a typical field-of-view size, or the primary beam size for radio aperture synthesis data. If one is looking for images containing a particular position, one should look for images with centers within a circle with diameter RoR, centered on the query position.

In the case of object catalogs RoR should normally be the largest of the typical size of the objects, the astrometric errors in the positions, or the resolution of the data. It may be some combination of these quantities when they are comparable. Users should note that catalogs may not be able to distinguish objects separated by an angular distance less than the RoR.

Generally providers may wish to specify an RoR conservatively, to ensure that a user querying the resource will get any data that might be relevant to the query but should be aware of the effects of the RoR on the volume of data returned to users.

*Coverage.Spectral* (string, list)
Definition: The spectral coverage of the resource.
Comment: Spectral coverage at the resource level will be in terms of general spectral regions (gamma-ray, x-ray, extreme UV, UV, optical, infrared, radio). The general spectral regions are defined specifically as follows:

| Coverage.Spectral | Represents |
| --- | --- |
| Radio | $\lambda \geq 10$ mm |
| | $\nu \leq 30$ GHz |
| Millimeter | $0.1$ mm $\leq \lambda \leq 10$ mm |
| | $3000$ GHz $\geq \nu \geq 30$ GHz |
| Infrared | $1\ \mu \leq \lambda \leq 100\ \mu$ |
| Optical | 0.3 μ ≤ λ ≤ 1 μ |
| | 300 nm ≤ λ ≤ 1000 nm |
| | 3000 Å ≤ λ ≤ 10000 Å |
| Ultraviolet | 0.01 μ ≤ λ ≤ 0.3 μ |
| | 100 Å ≤ λ ≤ 3000 Å |
| | 1.2 eV ≤ E ≤ 120 eV |
| X-ray | 0.1 Å ≤ λ ≤ 100 Å |
| | 0.12 keV ≤ E ≤ 120 keV |
| Gamma-ray | E ≥ 120 keV |

Resources containing data in multiple spectral regions may give a list (e.g., "Radio, Infrared").

*Coverage.Spectral.Bandpass* (string, list)
Definition: A specific bandpass specification.
Comment: Some resources and services may choose to give spectral coverage in more specific terms than the general spectral regions. The list of possible bandpass names is too lengthy to enumerate here, but would include optical bandpasses (U, V, B, R, I), narrow line filters (H-alpha, [OIII]), or other specific bandpass names.

*Coverage.Spectral.CentralWavelength* (float)
Definition: The central wavelength of the spectral bandpass, in meters.
Comment: This should be the most representative wavelength of the bandpass.

*Coverage.Spectral.MinimumWavelength* (float)
Definition: The minimum wavelength of the spectral bandpass, in meters.
Comment: Implementers are encouraged to set the minimum wavelength to be as inclusive as possible, allowing all relevant resources to be discovered.

*Coverage.Spectral.MaximumWavelength* (float)
Definition: The maximum wavelength of the spectral bandpass, in meters.
Comment: Implementers are encouraged to set the maximum wavelength to be as inclusive as possible, allowing all relevant resources to be discovered.

*Coverage.Temporal.StartTime* (string)
Definition: The earliest temporal coverage of the resource.
Comment: Temporal coverage specifications will be given in ISO8601 format.
An empty value field implies that there is no known earliest temporal coverage.

*Coverage.Temporal.StopTime* (string)
Definition: The latest temporal coverage of the resource.
Comment: Temporal coverage specifications will be given in ISO8601 format.
An empty value field implies that there is no known latest temporal coverage, i.e., that information continues to be added to the resource.

*Coverage.Depth* (float)
Definition: The (typical) depth coverage, or sensitivity, of the resource. Coverage.Depth is specified in flux density (Jy).

*Coverage.ObjectDensity* (float)
Definition: The (typical) density of objects, catalog entries, telescope pointings, etc., on the sky, in number per square degree.

*Coverage.ObjectCount* (int)
Definition: The total number of objects, catalog entries, telescope pointings, etc., in the resource.

*Coverage.SkyFraction* (float)
Definition: The fraction of the sky represented in the observations, ranging from 0 to 1.

*Resolution* (float)
Definition: The resolution of the resource contents.
Comment: Resolution is divided into the following sub-elements:

*Resolution.Spatial* (float)
Definition: The spatial (angular) resolution that is typical of the observations, in decimal degrees.

*Resolution.Spectral* (float)
Definition: The spectral resolution that is typical of the observations, given as the ratio $\lambda/\Delta\lambda$ (so that higher spectral resolution has a larger number).

*Resolution.Temporal* (float)
Definition: The temporal resolution that is typical of the observations, given in seconds.

*UCD* (string, list)
Definition: A list of the UCDs (Unified Content Descriptors, http://cdsweb.u-strasbg.fr/doc/UCD.htx) represented in the resource.
Comment: Some large or complex resources will have hundreds of associated UCDs and are unlikely to be specified in the resource metadata. Users of the resource metadata should not assume that an empty specification implies that the resource has no associated UCDs.

*Format* (string, list)    [Dublin Core]
Definition: The physical or digital manifestation of the information provided by the resource.
Comments: Typical values would be "image/fits", "image/gif", "text/plain", "text/html", "text/xml" (for VOTable), etc. MIME types should be used where available to specify digital information formats in order to utilize existing standards.

Other format values will be used to describe the physical medium of the information: CDROM, Digital Planetarium, Online, Presentation, Print, Slides, Video. Format specifications may be combined, as in "Video, video/mpeg" (both hardcopy video cassettes and on-line MPEG files) or "CDROM, image/fits, image/gif" (FITS and GIF images are available on-line and on CDROM hardcopy).

*Rights* (string)        [Dublin Core]
Definition: Information about rights held in and over the resource.

Comment:  Dublin Core uses Rights to describe copyright and other intellectual property rights issues.  In the VO context Rights would describe access privileges, using the following values: public, proprietary, mixed.

## 3.5  Correspondence of collection and service content metadata with the Space-Time Coordinate schema

The complete syntax for the spatial coverage specification is described in the [Space-Time Metadata definition document](http://www.ivoa.net/Documents/latest/STC.html) (http://www.ivoa.net/Documents/latest/STC.html) under ResourceProfile/AstroCoordArea.

Coverage.Spatial
    ResourceProfile/AstroCoordArea/SpatialInterval
        SpatialInterval may be PositionInterval or Region

Coverage.Spectral
    The "band" definitions do not exist; everything is explicitly numerical

Coverage.Spectral.Bandpass
    The bandpass definitions do not exist; everything is explicitly numerical

Coverage.Spectral.CentralWavelength
    ResourceProfile/AstroCoords/Spectral/Value

Coverage.Spectral.MinimumWavelength
    ResourceProfile/AstroCoordArea/SpectralInterval/LoLimit

Coverage.Spectral.MaximumWavelength
    ResourceProfile/AstroCoordArea/SpectralInterval/HiLimit

Coverage.Temporal.StartTime
    ResourceProfile/AstroCoordArea/TimeInterval/StartTime

Coverage.Temporal.StopTime
    ResourceProfile/AstroCoordArea/TimeInterval/StopTime

Coverage.SkyFraction
    ResourceProfile/AstroCoordArea/SpatialInterval(FillFactor)
    Caveat: it is only the fraction of the specified region

Resolution.Spatial
    ResourceProfile/AstroCoords/Position/Resolution

Resolution.Spectral
    ResourceProfile/AstroCoords/Spectral/Resolution

Resolution.Temporal
    ResourceProfile/AstroCoords/Time/Resolution

Uncertainty.Spatial

ResourceProfile/AstroCoords/Position/Error

Uncertainty.Spectral
ResourceProfile/AstroCoords/Spectral/Error

Uncertainty.Temporal
ResourceProfile/AstroCoords/Time/Error

RegionOfRegard
ResourceProfile/AstroCoords/Position/Size

# 4 Data and metadata quality assessment

Users of virtual observatory resources need some way to assess the quality of the data and of the associated descriptive information in the registry. Data quality is both subjective and quantitative, and data collections may have no single data quality metric. While the completeness and consistency of the resource metadata itself may be a reasonable indicator of the quality of the associated resource, this is at best a qualitative measure. The following metadata elements are intended to capture the most basic measures of data quality, and may well require extensions as VO usage practices evolve and become more sophisticated.

*DataQuality* (char)
Definition: An overall assessment of the integrity, consistency, and level of documentation concerning uncertainty estimates and calibration procedures, of the data provided by the resource. We suggest 3 general grade levels, plus codes for unknown or undocumented cases:

- A    Data are fully calibrated, fully documented, and suitable for professional research.
- B    Data are calibrated and documented, but calibration quality is inconsistent. Users are advised to check data carefully and recalibrate.
- C    Data are uncalibrated.
- U    Data quality is unknown. If a resource does not provide a data quality assessment, class U should be assumed.

*ResourceValidationLevel* (int)
Definition: A numeric grade describing the quality of the resource description and interface, when applicable, to be used to indicate the confidence an end-user can put in the resource as part of a VO application or research study. The allowed values are:

- 0    The resource has a description that is stored in a registry. This level does not imply a compliant description.

- 1    In addition to meeting the level 0 definition, the resource description conforms syntactically to this standard and to the encoding scheme used.

- 2    In addition to meeting the level 1 definition, the resource description refers to an existing resource that has been demonstrated to be functionally compliant.

When the resource is a service, it is considered to exist and to be functionally compliant if use of the Service.InterfaceURL or Service.BaseURL responds without error when used as intended by the resource. If the service is a standard one, it must also demonstrate the response is syntactically compliant with the service standard in order to be considered functionally compliant. If the resource is not a service, then the ReferenceURL must be shown to return a document without error.

3   In addition to meeting the level 2 definition, the resource description has been inspected by a human and judged to comply semantically to this standard as well as meeting any additional minimum quality criteria (e.g., providing values for important but non-required metadata) set by the human inspector (see comment below).

4   In addition to meeting the level 3 definition, the resource description meets additional quality criteria set by the human inspector and is therefore considered an excellent description of the resource. Consequently, the resource is expected to be operate well as part of a VO application or research study.

If no value is provided, level 0 should be assumed.

Comment: Unlike other resource metadata, ResourceValidationLevel values will most often be set by entities other than the resource provider. The most common assigners of this metadatum will be registry administrators. In an environment where a resource record may exist in many registries, each instance may have a different value given, depending on the practice and quality standards set by the registry. Levels 0, 1, and 2 are defined so that they can be assigned to resource descriptions automatically by a software agent (e.g. operating on behalf of a registry). Levels 3 and 4, by definition, require human review. Curators of resource descriptions (such as registry administrators) who conduct human inspection should base their own quality criteria first on the comments given in this document. It is recommended that the level 4 criteria include the requirement that the description contain a legal value for the DataQuality metadatum.

*ResourceValidatedBy* (URI)
Definition: The IVOA identifier for the registry or organisation that set the value of ResourceValidationLevel.

*Uncertainty.Photometric* (float)
Definition: The uncertainty of the photometric measurements provided by the resource, given in Jy.

*Uncertainty.Spatial* (float)
Definition: The uncertainty of the astrometric, or positional measurements, provided by the resource, given in degrees.

*Uncertainty.Spectral* (float)
Definition: The uncertainty of the wavelengths provided by the resource, given in meters.

*Uncertainty.Temporal* (float)

Definition: The uncertainty of the temporal measurements provided by the resource, given in seconds.

# 5 Service metadata concepts

The metadata necessary for describing a service will vary quite a bit depending on the type of service it is. We propose two general categories of service metadata:

*Interface metadata*, which describe how to access the service—the inputs and the outputs. There will be standard types of interfaces that could include a web-browser-based interface (i.e., HTML Forms), a Web Service interface (describable by a WSDL document), a general HTTP Get interface (e.g., using *key=value* arguments), and a GLU-described interface.

*Capability metadata*, which describe what the service does, its limitations, and other behavioral characteristics.

Note that these categories are reasonably orthogonal. We can imagine the same basic service—in terms of its capabilities—accessible through multiple interfaces.

We expect that for each standard service recognized by the VO there will be a specification document that defines all the specific metadata necessary to describe a particular implementation of that service; thus, we do not include them all here. However, we can identify a few metadata concepts that might be employed to describe a particular service. Described below, these concepts should be employed by standard service specifications wherever they are applicable. We note also that metadata associated with the VOTable schema can also be reused to describe the inputs and outputs of a service that returns a VOTable.

## 5.1 Interface metadata

*Service.AccessURL* (URL)
Definition: The URL that a client uses to access a service.
Comment: The Service.AccessURL is the endpoint URL for the particular service. How this URL is to be interpreted and used depends on the particular type of service (e.g., REST vs. web service). If such interpretation and usage cannot be determined based on the service type, or if a service does not use standard defaults, Service.DefinitionURL and Service.BaseURL may be used for clarification.

*Service.DefinitionURL* (URL)
Definition: A URL pointing to a document that presents or describes the service interface.
Comment: For a Web Service, this would point to the WSDL document, for a GLU-described service, it would point to the GLU record, and for a browser-based service, this would be the Web page that actually contains the Web Form.

*Service.BaseURL* (URL)

Definition: The base portion of a URL used to invoke a service with the expectation that an additional string must be appended for the service to execute properly. The syntax of the appended string is defined by the specific service.

*Service.HTTPResultsMIMEType* (MIME type)
Definition: The MIME type that is returned by a service.

## 5.2 Capabilities metadata

*Service.StandardID* (URI)
Definition: An identifier for a standard service. The syntax for Identifiers is described in *IVOA Identifiers* in the IVOA document collection (http://www.ivoa.net/Documents/).
Comment: This provides a unique way to refer to a service specification standard, such as a Simple Image Access service. It assumes that such standard is registered and accessible. The Service.StandardID should include a specification of the version number or release number of the given service.

*Service.MaxSearchRadius* (float, decimal degrees)
Definition: Service providers may choose to restrict the scope of searches done against their services, lest they be swamped with requests for millions or billions of results records. Service.MaxSearchRadius restricts searches to some maximum radius (in decimal degrees) about a celestial coordinate.
Comment: A value of 180.0 or greater denotes that there is no restriction.

*Service.MaxReturnRecords* (int)
Definition: Service providers may choose to restrict the number of records returned in order to avoid swamping the user with responses to an overly general query. If no value is provided, it is assumed that there is no restriction on the number of records returned.
Comment: Some resources may be limited in the number of records available, e.g., there are only 257 objects in a particular catalog. Service.MaxReturnRecords should not be set to 257 in such a case. Service.MaxReturnRecords is intended to show a server-side constraint to requests, not the size or scale of the resource itself. The metadata element Coverage.ObjectCount should be used for this purpose.

*Service.MaxReturnSize* (float, bytes)
Definition: Service providers may choose to restrict the total size of a service response in order to manage computational resources. Service.MaxReturnSize allows providers to state such a restriction. The value is floating point in order to accommodate sizes larger than 2147483647 bytes (2 gigabytes). If no value is provided, it is assumed that there is no restriction on the size of the service response.

## 6 Example

Example: The Sloan Digital Sky Survey data as hosted by MAST at STScI (with no assertion that the metadata element values are actually correct, though they are not unreasonable).

Identity metadata
    Title                    Sloan Digital Sky Survey

| | | |
|---|---|---|
| | ShortName | SDSS |
| | Identifier | ivo://stsci.edu/mast/sdss |

Curation metadata
| | | |
|---|---|---|
| | Publisher | Space Telescope Science Institute/MAST |
| | PublisherID | ivo://stsci.edu/mast |
| | Creator | Sloan Digital Sky Survey Consortium |
| | Creator.Logo | http://archive.stsci.edu/images/sdss_logo.gif |
| | Contributor | Sloan Digital Sky Survey Consortium |
| | Date | 2003-02-01 |
| | Version | SDSS EDR |
| | Contact.Name | Archive Branch, Space Telescope Science Institute |
| | Contact.Address | 3700 San Martin Drive, Baltimore, MD 21218 USA |
| | Contact.Email | archive@stsci.edu |
| | Contact.Telephone | +1-410-338-4547 |

General content metadata
| | | |
|---|---|---|
| | Subject | galaxies, quasars, stars, CCD photometry, spectroscopy, redshift, sky surveys |
| | Description | The Sloan Digital Sky Survey is using a dedicated 2.5 m telescope and a large format CCD camera to obtain images of over 10,000 square degrees of high Galactic latitude sky in five broad bands (u', g', r', i' and z', centered at 3540, 4770, 6230, 7630, and 9130 Å, respectively). Medium resolution spectra will be obtained for approximately $10^6$ galaxies and 100,000 quasars. The early data release (EDR), on June 2001, includes searchable catalogs of images and spectra, images for display and scientific purpose in both 2-D FITS and JPEG formats, and spectra in both 1-D FITS and GIF formats. The EDR covers about 460 square degrees of sky. The next data releases will occur every 18 months or so. |
| | Source | 2002AJ….123..485S |
| | ReferenceURL | http://archive.stsci.edu/sdss/index.html |
| | Type | Survey, Catalog, EPOResource |
| | ContentLevel | Research |
| | Relationship | mirror-of |
| | RelationshipID | ivo://sdss.org/sdss/edr |

Collection and service content metadata
| | | |
|---|---|---|
| | Facility | Apache Point Observatory, Sloan 2.5-m Telescope |
| | Instrument | Five-band clocked CCD camera |
| | Coverage.Spatial | PositionInterval FK5 145.17 –1.25 235.9 1.25 PositionInterval FK5 250.71 52.15 267.0 66.29 PositionInterval FK5 350.43 –1.25 359.99 1.17 PositionInterval 0.0 –1.25 56.37 1.17 |
| | Coverage.RegionOfRegard | 0.0001 |
| | Coverage.Spectral | Optical |
| | Coverage.Spectral.Bandpass | u', g', r', i', z' |
| | Coverage.Spectral.MinimumWavelength | 400.e-9 |
| | Coverage.Spectral.MaximumWavelength | 850.e-9 |
| | Coverage.Temporal.StartTime | 1999-12-25 |
| | Coverage.Temporal.StopTime | 2001-07-15 |
| | Coverage.Depth | 3.e-6 |
| | Coverage.ObjectDensity | 6.e4 |
| | Coverage.ObjectCount | 2.e7 |
| | Coverage.SkyFraction | 0.01 |

```
Resolution.Spatial              0.00028
Resolution.Spectral             5000
Resolution.Temporal             120
UCD             Not Provided
Format          text/xml
Rights          Public
```

Data quality metadata
```
DataQuality             A
ResourceValidationLevel 4       [provided by registry curator]
ResourceValidatedBy     ivo:/us-vo.org/registry
Uncertainty.Photometric 3.e-7
Uncertainty.Spatial     0.00003
Uncertainty.Spectral    1.e-11
Uncertainty.Temporal    0.1
```

Service metadata
```
Service.AccessURL               http://archive.stsci.edu/cgi-bin/sdss/catalog
Service.InterfaceURL            http://archive.stsci.edu/sdss/catalog.html
Service.BaseURL                 http://archive.stsci.edu/cgi-bin/sdss/catalog
Service.HTTPResultsMIMEType     text/xml
Service.StandardID              ivo://ivoa.net/Services/ConeSearch

Service.MaxSearchRadius         0.2
Service.MaxReturnRecords        5000
Service.MaxReturnSize           5.e8
```

# 7  Changes from previous versions

From V1.01 to V1.1
- Expanded definitions of Title and ShortName (Section 3.1).
- Expanded definitions of Creator and Contributor (Section 3.2).
- Expanded definition of Date and changed to a required element (Section 3.2).
- Added Contact.Address and Contact.Telephone, and modified description of Contact to reflect these new elements (Section 3.2).  Updated example (Section 6).  (These elements have been added, deleted, and now added back again!).
- Expanded definitions of Subject and Description (Section 3.3).
- Expanded definition of Facility (Section 3.4).
- Redefined syntax for region specifications for Coverage.Spatial to be consistent with, and the most elementary version of, the Space-Time Coordinates specification. Eliminated Ellipse, Sector, and Constraint region specifications and added Position-Interval and AllSky region specifications (Section 3.4).
- Redefined element Coverage.RegionOfRegard (Section 3.4).
- Added Section 3.5 to show correspondence between Coverage information and the Space-Time Coordinates schema.
- Changed title and overview paragraph of Section 4 to include quality of metadata.
- Added element ResourceValidationLevel  (Section 4).
- Added element ResourceValidatedBy (Section 4).
- Added element Service.AccessURL to reflect actual usage of service URLs in the registry schemas (Section 5.1).

- Renamed Service.InterfaceURL to Service.DefinitionURL to be consistent with registry metadata schemas (Section 5.1).
- Renamed Service.StandardURI to Service.StandardID, consistent with example and with other elements that have the form of URIs/Identifiers. This was supposed to have been done in the previous version, but was overlooked. (Section 5.2)
- Deleted element ServiceStandardURL (Section 5.2).
- Expanded definition of Service.MaxReturnRecords (Section 5.2).
- Added element Service.MaxReturnSize (Section 5.2).
- Updated the example to reflect the above changes (Section 6).
- Changed all spellings of "organization" to "organisation".
- Fixed minor typographical errors and some line-spacing problems.

From V1.0 to V1.01
- Revised wording in Section 2 to clarify phrase concerning "independence" of resource metadata.
- Added text to indicate that PublisherID must have the form of a valid IVOA Identifier (Section 3.2).
- Clarified definition of ReferenceURL (Section 3.3).
- Added text to indicate that RelationshipID must have the form of a valid IVOA Identifier (Section 3.3).
- Deleted Convex from list of possible values for Spatial.Coverage, as it is redundant with extant ability to specify region intersections (Section 3.4).
- Corrected URL to coordinate system descriptions at CDS (http://aladin.u-strasbg.fr/java/doctech/cds/astro/Astroframe.html, Section 3.4).
- Renamed Coverage.Spectral value of UV to Ultraviolet, and deleted the value EUV. EUV spectral range is merged into UV. (Section 3.4.)
- Renamed Service.HTTPResults to Service.HTTPResultsMIMEType (Section 5.1 and in example).
- Renamed Service.StandardURI to Service.StandardID, consistent with example and with other elements that have the form of URIs/Identifiers. This was an inconsistency not noticed by reviewers previously. Also added text to indicate that Service.StandardID must have the form of a valid IVOA Identifier. (Section 5.2)
- Corrected URL in the example for element Service.InterfaceURL.

From V0.82 to V1.0
- Removed Creator from list of required elements (Section 3.2).
- Removed Contact.Address and Contact.Telephone, and modified description of Contact to reflect the removal of these elements (Section 3.2).
- Updated the example in Section 6 to reflect the above changes.

From V0.81 to V0.82:
- Included full change history from previous versions.
- Identified a subset of metadata elements as *required* for all resources. These are Title, Identifier, Publisher, Creator, Subject, Description, ReferenceURL, and Type.
- Added reference to IVOA Identifiers specification under metadata element Identifier (Section 3.1).
- Added Contact.Address and Contact.Telephone, and modified description of Contact to reflect these new elements (Section 3.2).
- Added alternate sources of Subject keywords in addition to the IAU Thesaurus (Section 3.3).

- Clarified definition of Source to avoid confusion with Dublin Core sense of "resource" (Section 3.3).
- Added Registry to list of recognized resource Types (Section 3.3).
- Added Other to list of recognized resource Types (Section 3.3).
- Added Relationship and RelationshipID to provide means to show relationships (mirror of, service for, derived from) between resources (Section 3.3).
- Restructured Section 3.3 to distinguish the most generic content metadata (which remains in Section 3.3) from content metadata that applies more specifically to data collections and their corresponding data access services (new Section 3.4).
- Refined definition of Coverage.Spatial (Section 3.4).
- Inserted Millimeter bandpass definition into Coverage.Spectral (Section 3.4).
- Added UCD to Section 3.4 to provide a way to show the contents of resources via their associated Unified Content Descriptors.
- Updated the example in Section 6 to include new metadata elements and use Identifiers that are consistent with the most recent drafts of the Identifier specification.
- Added acknowledgements to funding agencies.

From V0.8 to V0.81:
- Restricted units of Coverage.Depth to Jy and eliminated Coverage.Depth.Units. It is assumed that a registry data entry interface could handle unit conversions.
- Added section on data quality metadata, including elements DataQuality, Uncertainty.Photometric, Uncertainty.Spatial, Uncertainty.Spectral, and Uncertainty.Temporal.
- Changed names of Service-related metadata elements to separate Service root from the rest of the name (e.g., ServiceInterfaceURL → Service.InterfaceURL).
- Changed name of ServiceMSR to Service.MaxSearchRadius.
- Added Service.MaxReturnRecords.
- Updated example to reflect above.

From V0.7 to V0.8:
- Title: Document renamed to "Resource Metadata for the Virtual Observatory" recognizing that services are a subset of resources.
- Section 2: Made several minor wording changes to reflect generic use of the term "resource", which encompasses "service".
- Section 2: Made definition of "Curation Metadata" more specific.
- Section 2: Added definitions for metadata elements that are not provided, unknown, or not applicable. Replaces previous discussion at end of Section 3.3.
- Section 2: Added definitions for aggregate resources whose contents might include "any" or "all" of a given metadata element's types.
- Sections 2 and 3: Added note in Section 2 describing relationship of VO metadata to Dublin Core. In Section 3, changed references to Dublin Core to cite those elements that are the same as DC rather than those elements that are different from DC.
- Section 3.1: Changed Ticker to ShortName and extended string length from 8 to 16 characters.
- Section 3.2: Changed "based on" to "drawn from" the IAU Astronomy Thesaurus, concerning the selection of Subject metadata.
- Section 3.2: Moved elements Title, Subject, and ReferenceURL from 3.2 to 3.3.
- Section 3.3: Added metadata element Source (from Dublin Core), allowing for a cross reference to the bibcode (or other form of citation) for an associated bibliographic reference.
- Section 3.3: Extended Type to include Organisation and Project.

- Section 3.3: Modified Coverage.Spatial to conform to Region specification of Space-Time Metadata definitions.
- Section 3.3: Clarified definition of RegionOfRegard (which is NOT the angular resolution).
- Section 3.3: Added Coverage.Depth.Units to provide separate field for units string.
- Section 3.3: Added Coverage.SkyFraction.
- Section 3.3: Added Resolution.Spatial, Resolution.Spectral, and Resolution.Temporal elements.
- Section 3.3: Allowed Facility and Instrument elements to be lists.
- Section 3.3: Changed examples for Format to use standard MIME types.
- Section 5: Updated example to include above changes and additions.

From V0.6 to V0.7:
- Reformatted document to be compatible with new IVOA documentation standards.
- Modified much of the architecture description to reflect current concepts about the function and structure of registries.
- Distinguished *identity metadata* from other curation metadata.
- Added Ticker element to provide shorthand abbreviation for resource Title.
- Added PublisherID element to provide unique identifier for the resource publisher.
- Replaced *ResourceURL* with *ReferenceURL*.
- Deleted *ServiceURL* and incorporated more general service metadata (Section 4).
- Incorporated more complete metadata definitions for Education and Public Outreach resources. Affects: Type, Format, ContentLevel.
- Added Type = Simulation for theoretical models.
- Coverage.Spatial refers to Space-Time Metadata document for spatial coverage specifications.
- Added Coverage.Spectral.CentralWavelength, Coverage.Spectral.MinimumWavelength, and Coverage.Spectral.MaximumWavelength to more fully describe the spectral bandpass.
- Replaced Coverage.Temporal with Coverage.Temporal.StartTime and Coverage.Temporal.StopTime to be consistent with Date and use of ISO 8601 time format.
- Added Coverage.RegionOfRegard to aid users of the registry in assessing the utility of a resource.
- Added Coverage. Depth, Coverage.Object Density, and Coverage.ObjectCount to provide information on resource depth of coverage (sensitivity) and richness.
- Broadened definition of Format, consistent with Dublin Core, to include specifications of both physical and digital formats.
- Added new section on service metadata, introducing metadata elements ServiceInterfaceURL, ServiceBaseURL, ServiceHTTPResults, ServiceStandardID, ServiceStandardURL, and ServiceMSR.
- Updated example to use new metadata elements.

From V0.5 to V0.6:
- Added *Date* element, following Dublin Core and in response to October 2002 discussion.
- Added *Version* element, in response to October 2002 discussion. Not in Dublin Core.
- Added *Creator.Logo* element, in response to October 2002 discussion. Not Dublin Core.
- Added *Subject* element, to replace *Coverage.Topics*, as the former is Dublin Core, the latter is not, and the intent is basically the same.

- Added *ContentLevel* element, to replace *Coverage.Level*. Upon reflection I thought that *ContentLevel* was both a better label and was of a different nature than the other *Coverage* elements. M. Voit will provide further inputs on appropriate element values.
- Updated example to use new metadata elements.